\documentclass{article}
\usepackage{graphicx} % Required for inserting images
\usepackage{amsmath,amssymb}
\usepackage{xcolor} %
\usepackage{tikz}
\usepackage{authblk}

\usepackage{graphicx}

\usetikzlibrary{arrows.meta}
\title{A Kalman Filter Based Approach to NV Diamond Data Fusion For Improved Temperature Sensing}
\author[1]{Shraddha Rajpal} 
\author[2]{Qiaochu Guo} 
\author[2]{Brendon A. McCullian} 
\author[3]{Tyrus Berry} 
\author[2]{Zeeshan Ahmedthanks{Corresponding author: zeeshan.ahmed@nist.gov}} % Added [3] here (or [4] if different)
\affil[1]{School of Mathematical and Statistical Sciences, Clemson University, Clemson, SC 29634, USA}
\affil[2]{National Institute of Standards and Technology, MML, Gaithersburg, MD, 20899, USA}
\affil[3]{Department of Mathematical Sciences, George Mason University, Fairfax, VA 22032, USA}

\date{March 2026}

\begin{document}
    
\maketitle
\section{Abstract}

Nitrogen‑vacancy (NV) centers in diamond have been demonstrated to enable high sensitivity temperature measurements using multiple modalities. Standalone optically detected magnetic resonance (ODMR) offers robust temperature estimates albeit with high latency, whereas all-optical based measurements provide millisecond resolution but suffer from poorer long-term accuracy. In this work we demonstrate a hot-start Kalman filtering approach that fuses the two modalities, leading to a 57\% improvement in accuracy. The fused estimate  achieves higher long-term accuracy with lower latency, demonstrating a viable route to implementation of self‑correcting, high‑precision NV‑diamond temperature sensing schemes.\\
\vspace{1em} \\
\noindent \textbf{Keywords:} NV diamond, Kalman Filter, data fusion, probabilistic model, Neural networks

\section{Introduction}

Thermometry plays a crucial role in modern technology, with applications spanning from industrial processing to biomedical research\cite{dedyulin2022emerging,machin_review}. The demand for accurate temperature measurement using novel technologies such as photonic thermometry has grown with increasing use of sensor networks for process control\cite{sensornetworks2025} and development of novel technologies such as quantum computing and quantum sensing\cite{dedyulin2022emerging,spie_ahmed}. Traditional thermometry techniques fall short in meeting the requirements of these emerging fields due to limitations in accuracy over long-time periods and operational conditions \cite{dedyulin2022emerging}. 
\par The size of the overall temperature sensor market is a powerful motivating force driving the development of novel technologies targeted towards meeting present and future measurement needs of the economy \cite{dedyulin2022emerging}. The existing metrology infrastructure and user expectations of minimum uncertainty metrics (10 mK to 1 K) along with   size, weight, power and cost (SWaP-C) requirements represent a significant barrier that any new temperature measurement technology \cite{AHMED_IMEKO} must overcome.  Over the past two decades, photonic thermometry has emerged as a potential competitor to legacy technologies. Considerable work has been done to explore novel materials (e.g. silicon\cite{xu2014ultra} and  silicon nitride\cite{Silicon_nitride}), device configurations (Bragg waveguides\cite{WBG_optics_letter, Yije_Fano}, ring resonators\cite{xu2014ultra,Krenek_ring_resonator, Sergey_ring}, photonic crystal cavities\cite{PhCC} etc) and instrumentation\cite{Krenek_ring_resonator, Fleisher:23} and modeling \cite{AHMED2022113872} intended to widen the application window of photonic thermometers beyond metrology labs. 
\par Nitrogen vacancy (NV) color centers in diamond have garnered significant attention for their potential in nanoscale thermometry, owing to their high sensitivity, stability, multi-modal sensing capabilities and bio-compatibility. These sensors leverage the sensitivity of NV center's electron spin and lattice changes  to environmental magnetic and electrical fields  \cite{maze2008nanoscale, dolde2011electric} along with sensitivity to local temperature and pressure to enable high sensitivity measurements. NV center has attracted considerable attention for applications in quantum magnetometry  \cite{maze2008nanoscale}. More recently its sensitivity to temperature  has led to efforts to develop NV center's temperature sensors for sensing in embedded systems \cite{neumann2013high, PNAS_10mk} and biomicroscopy applications\cite{kucsko2013nanometre}.  Temperature sensitivities on the order of $10\,\mathrm{mK}/\sqrt{\mathrm{Hz}}$ and $2\,\mathrm{K}/\sqrt{\mathrm{Hz}}$  have been reported using pulsed-ODMR and continuous-wave ODMR measurements, respectively \cite{Fujiwara_2021}. 

\par Recently, NV center's temperature measurements have taken a significant step forward, in particular  the application of machine learning (ML) to NV center based measurements has resulted in significant advances. Rajpal et al reported accuracies of  $\approx\pm  2$K for optical-fiber coupled ODMR-based NV center temperature sensors using physics-based and data-driven machine learning models \cite{rajpal2025evaluating}.  Yamamoto \emph{et al} \cite{Yamamoto_2025} have reported accuracies of $<$1K for confocal ODMR measurement using Gaussian Process regression to infer temperature directly from the ODMR spectra. These results have highlighted the potential of ML to advance the measurement science potential of ODMR-based measurements. 
\par A principal advantage of the ODMR based measurement scheme is that common causes of calibration drift such as lineshape changes due to changes in local microwave field strength and laser power can be compensated for using in-situ measurement schemes such as  Rabi oscillation measurement. Furthermore, since the ODMR spectra can be computed from first principles, ML models that are trained and/or constrained by quantum properties of the NV center can be deployed to enable a computational path to drift detection and eventual self-correcting sensors. A significant barrier to adoption of  such a physics-based self-correcting system is the measurement latency- limited by data acquisition and data processing times- that can range from hundreds of seconds to tens of minutes.  
\par In a recent study using fiber-coupled NV center photoluminescence (PL) sensors, we demonstrated uncertainties from 800 mK down to 15 mK and time resolution of 1 ms over a temperature range of $-40.0 ^{\circ}\mathrm{C}$ to $70.0 ^{\circ}\mathrm{C}$ using highly optimized deep neural networks \cite{rajpal_PL}. Furthermore, we developed a methodology, explainable modes of deviation (xMoDs), to provide a spectroscopist-friendly lens for interpreting the hidden layer activations of the neural network. This allows us to interpret the decision making of the neural network using our spectroscopic understanding of the NV center's PL. The NV center all-optical technique achieves millisecond time resolution, combined with superior accuracy and low SWAP-C ,making it well-suited for routine measurements on an industrial scale. However, unlike ODMR measurements, the all-optical technique is not currently amenable to a high accuracy physics-based modeling approach.  
\par A potential route to enable physics-based drift correction in NV center temperature sensors is to fuse ODMR and photoluminescence (PL) measurements using Kalman filtering. In such a scheme, PL measurements are used to make fast, accurate measurements and occasional time consuming ODMR measurements and associated computational workload is utilized to provide validation input to the corrector algorithm. By fusing PL and ODMR measurements, Kalman filtering can leverage the strengths of both techniques to achieve improved accuracy and reliability without sacrificing time resolution. 

\section{Experimental}
The experimental setup described below is used to perform %asynschronous continuous wave ODMR and
PL measurements on a fiber-coupled sensor.% to validate the persistence model described in section \ref{temporal}. %The ODMR sensor (Sensor A) is fabricated by affixing a $\approx 150 \ \mu \text{m}$ diameter diamond particle with $\approx $3.5$ \, \mu\text{mol/mol}$ of NV color centers (Adamas Nanotechnologies ) using \(\approx 1 \, \text{µL}\) \cite{equipment}  of UV curable optically clear epoxy (EPOTEK 301 G) onto port 3 of a wide band circulator (Thorlabs). A $50 \ \mu \text{m}$ diameter radio frequency (RF) wire is epoxied ontop of the micro-diamond. Port 1 of the wide band circulator is connected to a fiber-coupled 514 nm laser (LABS electronics p/n: DLnsec) whilst port 3 delivers the light to detector assembly, where long and short pass filters are used to selectively transmit light between 633 nm and 800 nm only. The filtered light is passed onto a 10X objective that focuses the light into a multimode $105 \ \mu \text{m}$ diameter optical fiber. The light is then carried to a photon counting module (Laser Components, p/n: COUNT-500). The photon detector's output is sent to a data acquisition card (National Instruments, NI-6363) via a coaxial cable. The PL sensor (Sensor B) is similarly fabricated by affixing a NV diamond sample, $\approx 150 \ \mu \text{m}$ diameter diamond particle with $\approx $3.5$ \, \mu\text{mol/mol}$ 
The sensor is fabricated by affixing a NV diamond sample, $\approx 150 \ \mu \text{m}$ diameter diamond particle with $\approx $3.5$ \, \mu\text{mol/mol}$ of NV color centers (Adamas Nanotechnologies\cite{equipment}) , using \(\approx 1 \, \text{µL}\) of UV curable optically clear epoxy ( EPOTEK 301 G) onto the sample end of 3-port fiber bundle reflection  probe.  The light port of the fiber probe is connected to the fiber-coupled laser (LABS electronics p/n: DLnsec) whilst the spectrometer leg of the bundle delivers the light to a single grating spectrometer (Princeton Instruments) that delivers the wavelength resolved PL to a liquid nitrogen cooled charged coupled device (CCD). The sensor is inserted in to a metrology grade drywell whose temperature was varied between $-40.0 ^{\circ}\mathrm{C}$ to $70.0 ^{\circ}\mathrm{C}$ . Drywell temperature was monitored using calibrated thermistor.
%The ODMR and PL sensors were asynchronously inserted in to a metrology grade drywell whose temperature was varied between 233 K and 343 K. Drywell temperature was monitored using calibrated thermistor. 
\par Synchronous ODMR and PL measurements for multirate data fusion described in section \ref{fusion} were carried out using a single fiber-coupled sensor over the temperature range of $15.0 ^{\,\circ}\mathrm{C}$ to $45.0 ^{\,\circ}\mathrm{C}$.  The sensor is fabricated by affixing a $\approx 150 \ \mu \text{m}$ diameter diamond particle with $\approx $3.5$ \, \mu\text{mol/mol}$ of NV color centers (Adamas Nanotechnologies) using \(\approx 1 \, \text{µL}\) \cite{equipment}  of UV curable optically clear epoxy (EPOTEK 301 G) onto port 3 of a wide band circulator. A $50 \ \mu \text{m}$ diameter radio frequency (RF) wire is epoxied ontop of the micro-diamond. Following assembly, the sensor is placed inside the drywell and connected to the laser. The collected scattered light was filtered using a single  650 nm longpass filter. For ODMR measurement the collected PL was directed to a  a photodiode. The photodiode ouptut current was trans-impedance amplified at a gain of 100 nA/V and fed into a lock-in amplifier. The microwaves used for ODMR measurements were amplitude modulated at 467 Hz. The resulting lock-in voltage referenced to this modulation frequency was recorded using a data acquisition card. For PL spectrum measurements, the collected PL light is directed to a handheld OceanView spectrometer (OCEAN-HDX-XR) operating at room temperature. 

%We pattern a microstrip on a flexible PCB (OSH Park Flex PCB) as the microwave antenna for NV-center excitation. NV-rich microdiamonds are sparsely affixed along the microstrip with epoxy. To control the sample temperature, a flexible resistive foil heater (Thorlabs HT10K) is attached to the back of the PCB. We use a homebuilt confocal microscope to obtain ODMR and PL spectra from the NV microdiamonds at different temperature. PL from the microdiamonds is sent through an optical fiber to either an OceanView spectrometer for PL spectrum collection or to a single photon counter (Excelitas SPCM-AQR-16-FC) for ODMR measurements. At each temperature, one ODMR spectrum and 10,000 PL spectra are collected back-to-back. 

\section{Methods: }

\subsection{Kalman Filter for Data Fusion}

We first recall the Kalman update equations for a linear system with state variable $x_i$ and observations $y_i$ indexed by time, 
\begin{align} x_{i+1} &= Fx_i + \Gamma \omega_i \\
y_{i+1} &=  Hx_{i+1} + J\nu_{i+1}
\end{align}
with independent mean-zero Gaussian dynamical/process noise $\omega_i \sim \mathcal{N}(0,Q)$ and observation noise $\nu_i \sim \mathcal{N}(0,R)$ (see for example \cite{simon}).  Given the posterior estimate of the state $x_{i-1}^a$ and covariance $P_{i-1}^a$ at step $i-1$, the Kalman update for a linear system is
\begin{align}
x_i^f &= Fx_{i-1}^a \\
y_i^f &= Hx_i^f \\
P_i^f &= FP_{i-1}^aF^\top + \Gamma Q \Gamma^\top \label{prior} \\
K_i &= P_i^f H^\top(HP_i^f H^\top + JRJ^\top)^{-1} \label{gain} \\
x_i^a &= x_i^f + K_i(y_i - y_i^f) \\
P_i^a &= P_i^f - K_i HP_i^f \label{posterior}
\end{align}
where $x_i^f$ represents the forecast of the state given only the observations up to time $i-1$ and $P_i^f$ represents the covariance of the forecast.  Similarly, $y_i^f$ represents the forecast of the $i$-th observation given only the observations up to time $i-1$, and the difference between the true observation and the forecast, 
\[ \epsilon_i \equiv y_i - y_i^f  \] 
is called the \emph{innovation}.  The Kalman gain matrix $K_i$ optimally combines the forecast $x_i^f$ with the innovation to form the posterior estimate $x_i^a$ which is the maximal likelihood and minimum variance estimator of the true state $x_i$.  The filter also produces the covariance matrix $P_i^a$ of the estimator for use in the next filter step.

In our implementation, temperature observation sequences are first generated by three predictive models (autoencoder, convolutional neural network (CNN), and multilayer perceptron (MLP) based regressors) on held-out PL data. We note that the model predictions are 
paired with original sample index (proxy for time of measurement) to preserve temporal order, tuples \((\hat y_i, y_i^{\mathrm{ref}}, i)\) are saved for filtering. Kalman filtering is then performed in a separate stage using a scalar random-walk model (\(F=1\), \(H=1\)), with process and measurement uncertainty represented by \(Q\) and \(R\). By assuming the signal is slowly varying in time and that errors are not time correlated, this approach yields temporally smoothed temperature estimates while accounting for both process variability and measurement noise.

\subsubsection{Data Fusion}

Now we consider the case of multiple sensors, where sensors could be different modalities of single sensor or redundant sensors reporting the same modality.  We assume that each sensor is already combined with a technique for producing an estimate of the hidden variable, $x_i$, and we consider our full observation vector to be the concatenation of all of these estimates.  We denote by $x_i^1,...,x_i^m$ the estimates of $x_i$ produced by sensors $1,...,m$.  Each sensor will have its own independent noise sources, but it may also have noise sources which are correlated with those of the other sensors.  Thus, we write $x_i^j = x_i + \nu_i^j$ where $\nu_i^j$ is the noise at time $i$ impacting the $j$-th sensor.  We can now write our observation model as:
\[ y_i = \left[\begin{array}{c} x_i^1 \\ \vdots \\ x_i^m \end{array} \right] = \left[\begin{array}{c} x_i + \nu_i^1 \\ \vdots \\ x_i + \nu_i^m \end{array} \right] = \left[\begin{array}{c} 1 \\ \vdots \\ 1 \end{array} \right]x_i + \left[\begin{array}{c} \nu_i^1 \\ \vdots \\ \nu_i^m \end{array} \right] = Hx_i + \nu_i \]
where $H = [1,...,1]^T$ and $\nu_i = [\nu_i^1,...,\nu_i^m]^T$. While, the random noise variables in $\nu_i$ may not be Gaussian, we will make a Gaussian approximation to them, $\nu_i \sim \mathcal{N}(\overline \nu,R)$, by finding the mean, $\overline \nu$, which represents the bias of each sensor's estimate, and the joint covariance matrix, $R$.  These two statistics will be used by the Kalman filter to produce an unbiased, least squares estimate of $x_i$ which incorporates the information of each sensor together with information from past observations which the Kalman Filter has incorporated into the prior, given by $x_i^f$ and $P_i^f$.

\subsubsection{Temporal Prior \label{temporal}}

The Kalman Filter can outperform simple or weighted averaging of the various sensors' estimates because it incorporates a temporal prior which is typical of most sensors.  The most basic version of this prior is that at any given time, the most likely value for the hidden variable, $x_i$, is the prior value, $x_{i-1}$.  This assumption encodes a kind of continuity for the hidden variable in that observations close in time are expected to be correlated.  We can encode this prior by setting $F=I$ so that our filter uses the stationary prediction model:
\begin{equation}\label{stationarymodel} x_{i+1} = x_i + \omega_i. \end{equation}Since $\omega_i$ is assumed to have zero mean, the stationary model implies that the expected value of the hidden variable at the next time step is equal to its current value.  The stationary model variance parameter, $Q$, accounts for the dynamical noise building in a prior assumption regarding how fast the hidden variable changes.  %The stationary model also builds-in a prior assumption regarding how fast the hidden variable changes in the variance parameter, $Q$, for the dynamical noise.
If the hidden variable can change quickly, then relatively large changes are probable between time steps, which implies a large variance, $Q$.  Conversely, if the hidden variable changes slowly, then $Q$ should be small. The stationary model is commonly used for estimating parameters (a method known as \emph{state augmentation}) or anytime a model is not readily available \cite{friedland1969treatment,sitz2002estimation}. While the stationary model \eqref{stationarymodel} encodes the basic assumption of the hidden variable changing continuously in time, it is unable to capture trends or take advantage of additional smoothness in the temporal variations.  In such cases ,  autoregressive models can be incorporated to account for non-stationary behavior \cite{time_series_forecasting}.

\subsection{Estimating the Noise Statistics}

A Kalman filter works by combining different estimators with noise sources to obtain a better estimator via a form of weighted average.  In particular, it combines an observation with a prior estimate which summarizes all the information from previous observations combined.  In order to optimally combine these estimators, the filter requires accurate covariance matrices \cite{berry2013adaptive}. In particular the values $Q$ and $R$ which quantify the uncertainty in the system and observation noise respectively, are critical to the filter's performance\footnote{In general $Q$ and $R$ are matrices, however in this application the variables are scalar values and so $Q$ and $R$ are also scalars.}. When these values are unknown, adaptive filters can attempt to estimate them as part of the filtering procedure, however, it can take many filter cycles to obtain accurate values for $Q$ and $R$, and having good initial estimates is still critical to overall performance.

Obtaining an estimate of $R$ is straightforward when sensor calibration data is available.  Recall that $R$ is the covariance matrix of the observation noise and our observations are the temperatures predicted by a machine learning model.  As such the observation noise is simply the residual error between the temperature predicted by the machine learning model, and the true temperatures.  Since we are training the machine learning model before applying the Kalman filter, we assume that we have a training data set for which the true temperatures are available.  During training, we hold out a portion of the training data as a validation data set, which serves to ensure the trained machine learning model is generalizable to unseen data. After the machine learning model is fully trained (without seeing the validation set), we compute the residual errors on this validation set and the covariance of the residuals is used for $R$.  Note that this $R$ value will then be applied to the testing data which does not include either the training data nor the validation data.

Obtaining an estimate of $Q$ is more subtle since it quantifies the temporal uncertainty of the signal, also known as the \emph{system} or \emph{dynamical} noise.  The persistence model, employed in this study, tells the Kalman filter to assume that the best estimate for the true temperature at the current time is also the best initial estimate for the temperature at the next time.  This is a suitable assumption when temperature changes are significantly slower than the sensor response time. We note that the Kalman filter does not assume that the temperature cannot change, only that it is just as likely to increase as to decrease.  In order to combine its initial temperature estimate with the current observation at step $t_{i}$, the filter needs to know how much to expect the temperature to change during this time interval.  This can be very difficult to specify generically, since temperatures may change rapidly in some applications and environments and very slowly in others.  
\par The optimal $Q$ value would be the variance of the differences between the successive true temperature values.  While this can be computed for the training data set, it may or may not be indicative of the variations that will occur in the testing (out-of-sample) data, therefore some domain knowledge about the variability of the temperature is helpful in tuning $Q$.  Generally, $Q$ should scale roughly linearly with the time between observations since more time elapsed typically allows for larger changes in the underlying true temperature \cite{friedland1969treatment,friedland1982estimating}.  Here again, an adaptive filter can help alleviate these challenges by estimating and updating $Q$ as part of the filtering process \cite{berry2013adaptive}, but this may take many cycles to obtain the best results.  In order to demonstrate the capability of a well-tuned filter, we applied a simple grid search algorithm to find the best $Q$ value for the validation data. 
%, meaning that the Kalman filter was run on the validation data with various $Q$ values and the filtered estimates were compared to the true temperatures and the $Q$ value that produced the best estimates was chosen.  
This fixed $Q$ value was then applied to filter the testing sample data.

\subsection{Fusion of ODMR and PL Temperature Measurements\label{fusion}}

In this approach, we obtain temperature estimates from both the physics-constrained ODMR model and a data driven model for the PL data. The physics-constrained model has been previously described in detail \cite{rajpal2025evaluating}, briefly, we treat temperature calibration and estimation as a maximum likelihood estimation problem. We specify that ODMR spectra are well-described by using a bi-Lorentzian model whose peak centers are computed using quantum calculations and adjusted for temperature using empirical training data that fits higher order Taylor expansion of the zero-field term's temperature dependence. Once these terms are learned, we utilize these as deterministic parameters in an inverse model to estimate temperature given an input ODMR.

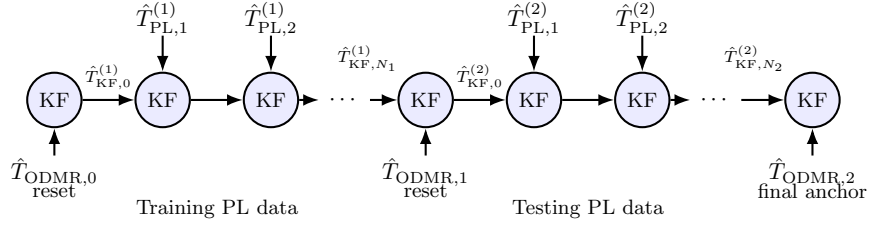
\begin{figure*}[th]
\centering
\resizebox{0.95\textwidth}{!}{%
\begin{tikzpicture}[
    x=1.18cm,
    y=0.95cm,
    kfnode/.style={
        circle,
        draw=black,
        fill=blue!8,
        minimum size=0.68cm,
        line width=0.9pt,
        font=\small
    },
    arr/.style={
        -{Latex[length=2.4mm]},
        line width=0.9pt,
        black
    },
    lab/.style={
        font=\normalsize,
        align=center
    },
    smalllab/.style={
        font=\scriptsize,
        align=center,
        fill=white,
        inner sep=1pt
    },
    notelab/.style={
        font=\small,
        align=center
    }
]

% KF nodes
\node[kfnode] (odmr0) at (1.0, 1.75) {KF};
\node[kfnode] (pl11)  at (2.4, 1.75) {KF};
\node[kfnode] (pl12)  at (3.8, 1.75) {KF};
\node[kfnode] (odmr1) at (5.8, 1.75) {KF};
\node[kfnode] (pl21)  at (7.2, 1.75) {KF};
\node[kfnode] (pl22)  at (8.6, 1.75) {KF};
\node[kfnode] (odmr2) at (10.8, 1.75) {KF};

% Horizontal KF arrows
\draw[arr] (odmr0.east) -- (pl11.west);
\node[smalllab] at (1.70, 2.12) {$\hat{T}_{\mathrm{KF},0}^{(1)}$};

\draw[arr] (pl11.east) -- (pl12.west);

\draw[arr] (pl12.east) -- (4.42, 1.75);
\node[lab] at (4.75, 1.75) {$\cdots$};
\draw[arr] (5.08, 1.75) -- (odmr1.west);
\node[smalllab] at (5.05, 2.42) {$\hat{T}_{\mathrm{KF},N_1}^{(1)}$};

\draw[arr] (odmr1.east) -- (pl21.west);
\node[smalllab] at (6.50, 2.12) {$\hat{T}_{\mathrm{KF},0}^{(2)}$};

\draw[arr] (pl21.east) -- (pl22.west);

\draw[arr] (pl22.east) -- (9.22, 1.75);
\node[lab] at (9.55, 1.75) {$\cdots$};
\draw[arr] (9.88, 1.75) -- (odmr2.west);
\node[smalllab] at (10.05, 2.42) {$\hat{T}_{\mathrm{KF},N_2}^{(2)}$};

% PL measurements from above
\node[lab] at (2.4, 3.05) {$\hat{T}_{\mathrm{PL},1}^{(1)}$};
\draw[arr] (2.4, 2.78) -- (pl11.north);

\node[lab] at (3.8, 3.05) {$\hat{T}_{\mathrm{PL},2}^{(1)}$};
\draw[arr] (3.8, 2.78) -- (pl12.north);

\node[lab] at (7.2, 3.05) {$\hat{T}_{\mathrm{PL},1}^{(2)}$};
\draw[arr] (7.2, 2.78) -- (pl21.north);

\node[lab] at (8.6, 3.05) {$\hat{T}_{\mathrm{PL},2}^{(2)}$};
\draw[arr] (8.6, 2.78) -- (pl22.north);

% ODMR measurements from below
\node[lab] at (1.0, 0.60) {$\hat{T}_{\mathrm{ODMR},0}$};
\draw[arr] (1.0, 0.86) -- (odmr0.south);
\node[notelab] at (1.0, 0.30) {reset};

\node[lab] at (5.8, 0.60) {$\hat{T}_{\mathrm{ODMR},1}$};
\draw[arr] (5.8, 0.86) -- (odmr1.south);
\node[notelab] at (5.8, 0.30) {reset};

\node[lab] at (10.8, 0.60) {$\hat{T}_{\mathrm{ODMR},2}$};
\draw[arr] (10.8, 0.86) -- (odmr2.south);
\node[notelab] at (10.8, 0.30) {final anchor};

% PL block labels
\node[notelab] at (3.15, 0.00) {Training PL data };
\node[notelab] at (7.90, 0.00) {Testing PL data};

\end{tikzpicture}%
}
\caption{KF scheme is visualized. The first ODMR estimate, labeled 0, resets the Kalman filter state, subsequently PL neural-network temperature estimates are assimilated sequentially within each PL block. The final ODMR estimate provides a terminal anchor.}
\label{fig:hot_start_kf_flowchart}
\end{figure*}

\par To estimate temperature from PL data, we first trained a neural network to estimate temperature from PL spectra. The PL dataset was organized by measured temperature. For each temperature, the 10,000 PL training data points were split so that the first 8,000 PL spectra were used for training the model and the next 2,000 spectra were used for validation. The PL spectra were standardized using only the training-set mean and standard deviation. A regularized fully connected ReLU neural network was then trained to predict temperature from the PL spectral features.

The neural network consisted of dense layers with sizes
$
512,\ 256,\ 128,\ 64,\ 32 .
$ Each dense layer was followed by batch normalization, rectified linear unit (ReLU) activation, and dropout. L2 regularization was applied to the dense-layer weights. The model was trained using the Adam optimizer with a learning rate of \(2 \times 10^{-4}\), a batch size of \(256\), early stopping based on validation loss, and learning-rate reduction when the validation loss plateaued.

After training, the PL neural network weights are frozen and model employed to predict temperature for all PL spectra. These predicted temperatures were treated as PL-derived temperature measurements for the Kalman filter. The PL measurement variance was estimated from the validation root mean squared error (RMSE) of the PL neural network:
$
R_{\mathrm{PL}} = \left(\mathrm{RMSE}_{\mathrm{PL,val}}\right)^2 .
$
Thus, the validation set was used to estimate the uncertainty of the PL temperature measurements. The testing data was not used to tune this value. For ODMR, three temperature estimates were available for each temperature condition. The first two ODMR repetitions were used to fit the parameters of the physics-informed ODMR spectral model, while the third repetition was reserved for testing. The ODMR measurement variance used in the Kalman filter was estimated from the training error of the ODMR estimator. Specifically, the ODMR training RMSE was squared and used as the initial ODMR covariance:
$
R_{\mathrm{ODMR}} = P_{\mathrm{ODMR,initial}} =
\left(\mathrm{RMSE}_{\mathrm{ODMR,train}}\right)^2 .
$ This value was used as the state covariance whenever the filter was initialized by an ODMR temperature estimate. The Kalman process noise for PL updates was fixed manually as
$
Q_{\mathrm{PL}} = 1 \times 10^{-4}.
$ This process noise determines how much uncertainty is added between consecutive PL observations. The Kalman filter was run separately for each temperature condition. At every available ODMR measurement, the filter was reinitialized using the ODMR-predicted temperature. Specifically, the filter state was reset as
\[
x = \hat{T}_{\mathrm{ODMR}},
\]
\[
P = R_{\mathrm{ODMR}},
\]
where \(x\) is the current fused temperature estimate, \(P\) is the uncertainty of that estimate, and \(\hat{T}_{\mathrm{ODMR}}\) is the ODMR-predicted temperature. After this ODMR initialization, the following PL neural-network temperature estimates were assimilated sequentially as noisy observations. For each PL observation \(z_{\mathrm{PL}}\), the filter performed the prediction and update steps:
\[
x_{\mathrm{pred}} = x,
\]
\[
P_{\mathrm{pred}} = P + Q_{\mathrm{PL}},
\]
\[
K = \frac{P_{\mathrm{pred}}}{P_{\mathrm{pred}} + R_{\mathrm{PL}}},
\]
\[
x = x_{\mathrm{pred}} + K\left(z_{\mathrm{PL}} - x_{\mathrm{pred}}\right),
\]
\[
P = (1-K)P_{\mathrm{pred}}.
\]
Here, \(K\) is the Kalman gain. It determines how much the fused estimate moves toward the new PL measurement. If the current uncertainty \(P_{\mathrm{pred}}\) is large relative to the PL measurement variance \(R_{\mathrm{PL}}\), the Kalman gain is larger and the filter gives more weight to the PL observation. If \(R_{\mathrm{PL}}\) is large, the filter trusts the PL measurement less.

\begin{figure*}[th]
    \centering
\includegraphics[width=0.51\linewidth]{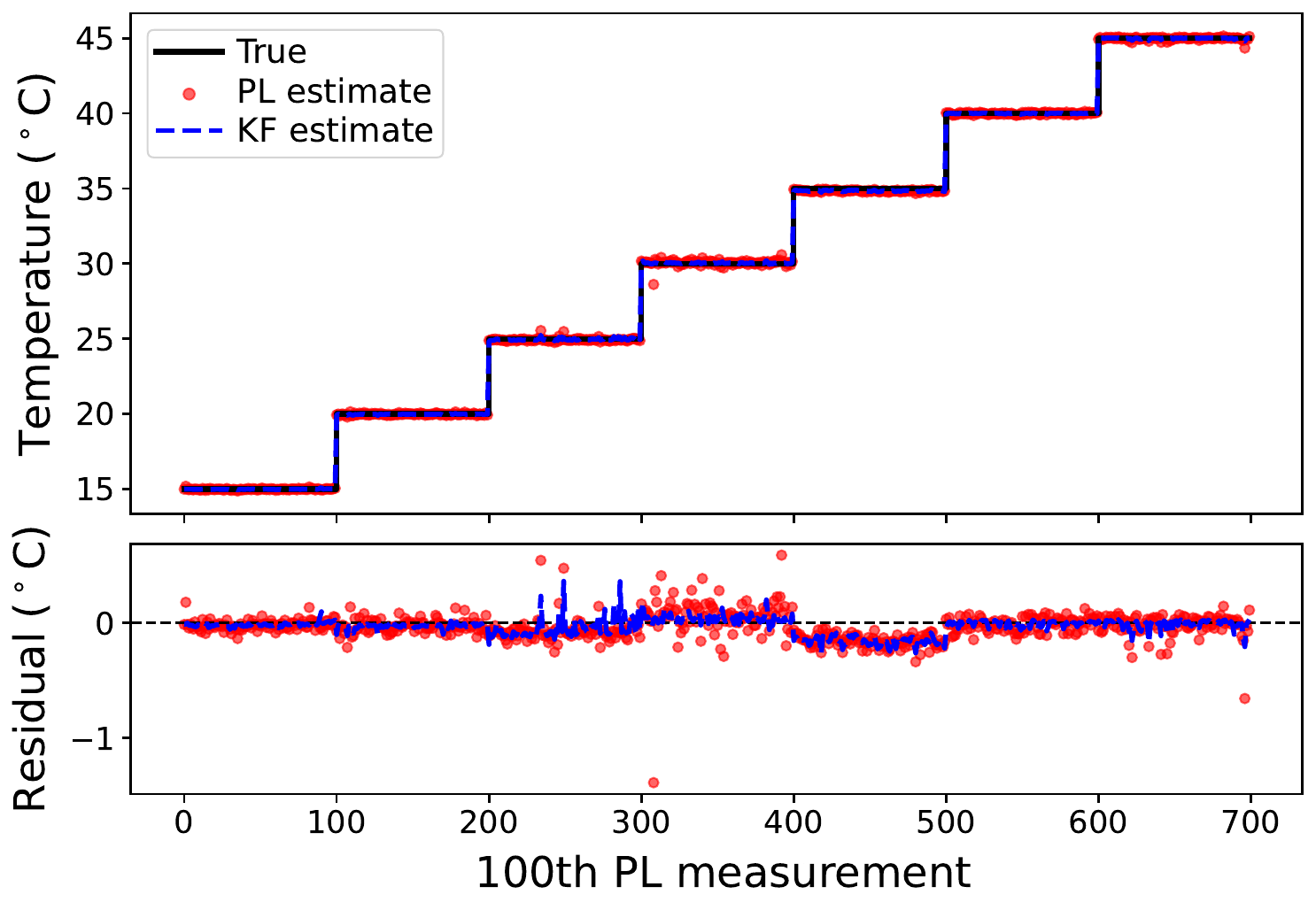}\includegraphics[width=0.51\linewidth]{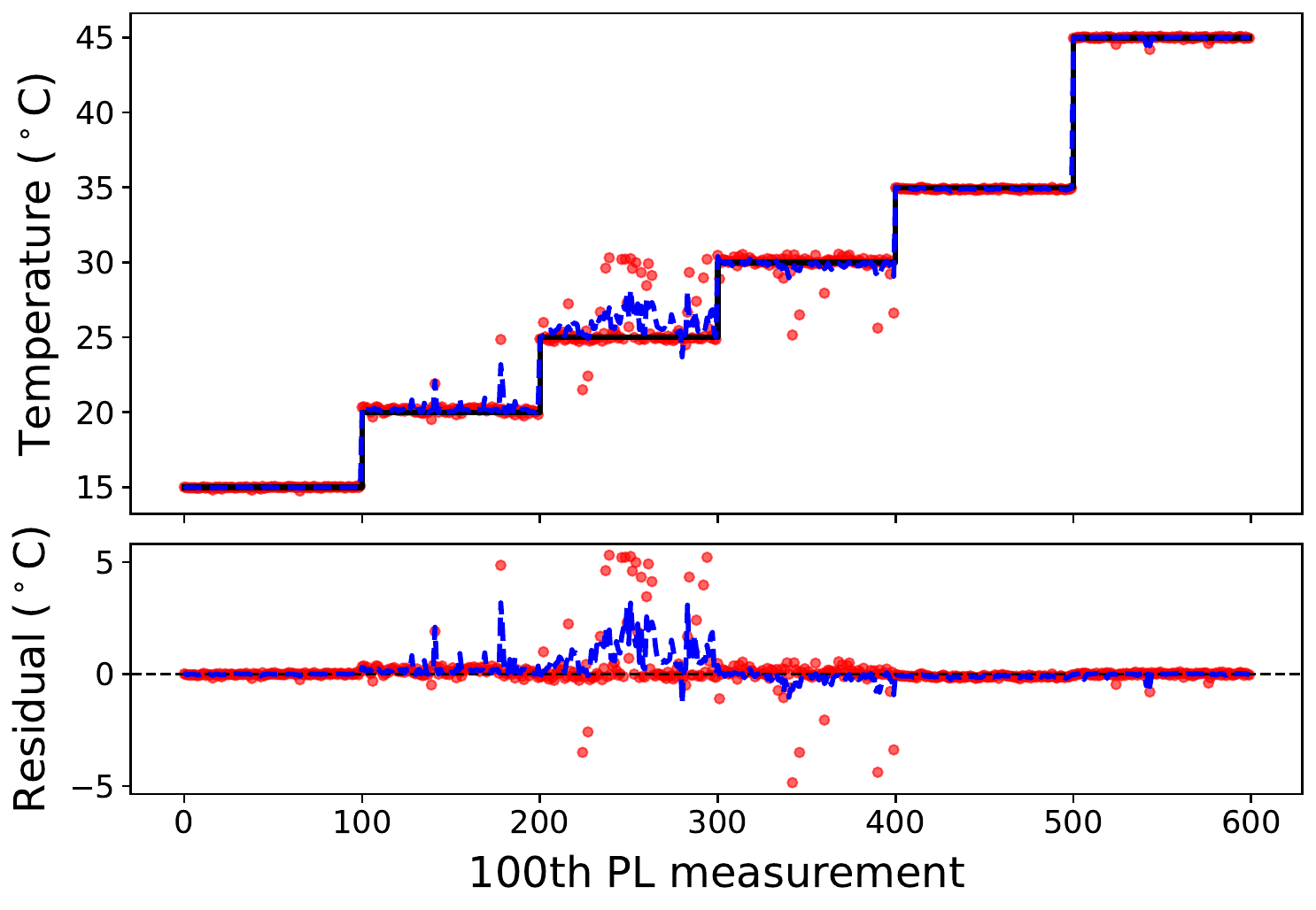}
    \caption{Left: For Training PL block shown is the true temperature, PL estimate estimate prior to KF, KF estimate, and their corresponding residuals. Right: Testing PL block showing the true temperature, PL estimate prior to the KF, KF estimate, and their corresponding residuals. Every 100th PL measurement is shown for visualization.}
    \label{fig:residuals}
\end{figure*}

The measurement sequence for each temperature condition, visualized in Fig \ref{fig:hot_start_kf_flowchart} was as following: The ODMR measurement is processed by the physics-informed model and a temperature estimate is produced. This first ODMR estimate is used to initialize the filter. Subsequently, training PL data  is assimilated sequentially. The arrival of the next ODMR measurement labeled, 1, reinitializes the filter, followed by sequential assimilation of Testing PL data. The last ODMR measurement in this sequence, labeled 2,  provides an anchor.
%\[\begin{aligned}
%&\text{ODMR repetition 0 initializes the filter,} \\
%&\text{Training PL data is assimilated sequentially,} \\
%&\text{ODMR repetition 1 reinitializes the filter,} \\
%&\text{PL block 2 is assimilated sequentially,} \\
%&\text{ODMR repetition 2 provides a final ODMR anchor.}
%\end{aligned}
%\]
We note that this method does not treat ODMR as a normal Kalman update blended with the previous state. Instead, each ODMR measurement acts as a hot start: it resets the fused estimate to the ODMR-predicted temperature and resets the uncertainty to the ODMR measurement variance. Thus, information accumulated from the preceding block of PL data  is not carried across an ODMR hot start.  The first block of PL data  was labeled as the training fusion block. It contained the first 10,000 PL spectra per temperature condition, which included the PL spectra used for training and validating the PL model. The second block of PL data was labeled as the testing fusion block. It contained the next 10,000 PL spectra per temperature condition when available. 
% In the current split, there were no out-of-sample PL spectra at \(40\,^{\circ}\mathrm{C}\), because the available \(40\,^{\circ}\mathrm{C}\) PL spectra were exhausted by the in-sample block.
The filter output was recorded after every ODMR reset and after every PL update. For each step, the measurement source, temperature condition, true temperature, raw measured temperature estimate, fused Kalman estimate, uncertainty \(P\), and Kalman gain \(K\) were stored.

Performance was evaluated by comparing two quantities against the known temperature labels:
\[
\mathrm{RMSE}_{\mathrm{Raw\ PL}}
=
\sqrt{
\frac{1}{N}
\sum_{i=1}^{N}
\left(
\hat{T}_{\mathrm{PL},i} - T_i
\right)^2
},
\]
\[
\mathrm{RMSE}_{\mathrm{\text\ KF}}
=
\sqrt{
\frac{1}{N}
\sum_{i=1}^{N}
\left(
\hat{T}_{\mathrm{KF},i} - T_i
\right)^2
}.
\]
Here, \(\hat{T}_{\mathrm{PL},i}\) is the raw PL neural-network temperature estimate, \(\hat{T}_{\mathrm{KF},i}\) is the Kalman-fused temperature estimate after PL assimilation, and \(T_i\) is the known temperature label.

The RMSE was reported separately for the training and testing PL blocks. The Kalman noise parameters were not tuned using the testing data. The process noise \(Q_{\mathrm{PL}}\) was fixed a priori, the PL measurement variance \(R_{\mathrm{PL}}\) was estimated from the PL validation error, and the ODMR variance \(R_{\mathrm{ODMR}}\) was estimated as the squared training RMSE of the ODMR temperature estimator. These values were then held fixed during testing evaluation.

\begin{figure*}
    \centering
\includegraphics[width=\linewidth]{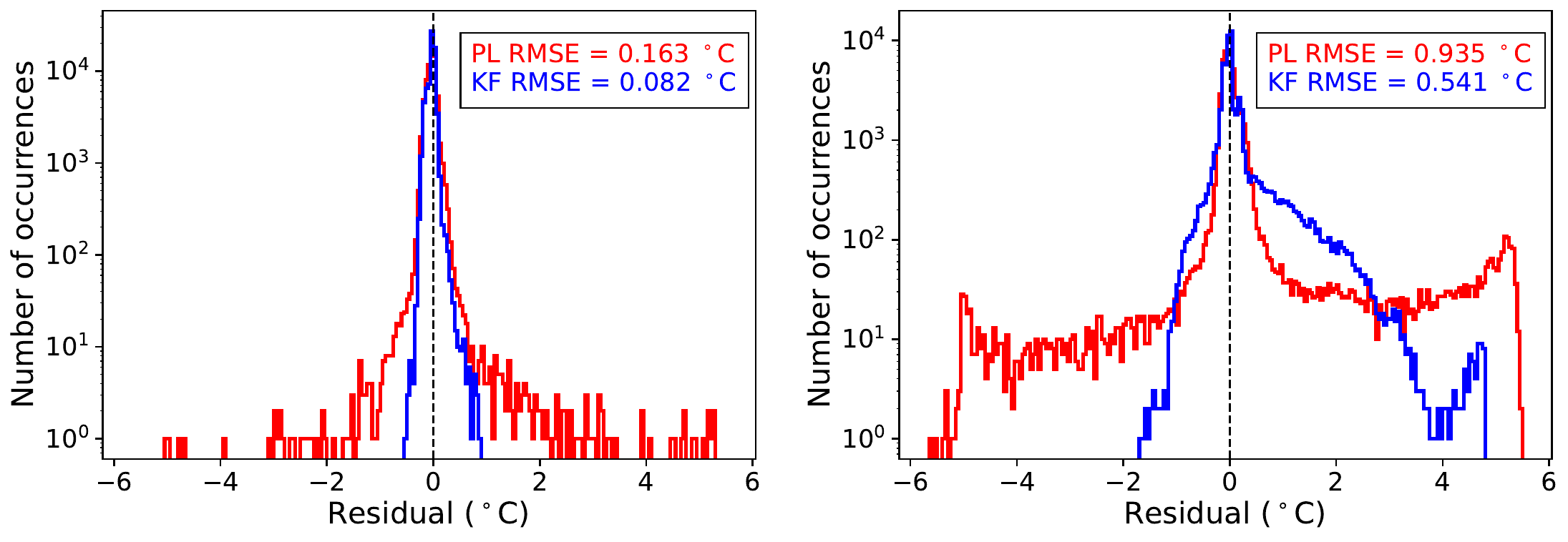}
\caption{Left: Training residual distribution for the PL and KF temperature estimates. Right: Testing residual distribution for the PL and KF temperature estimates.}
\label{fig:histograms}
\end{figure*}

\subsection{Results and Discussion}
As noted above, we evaluated temporal prior on the Kalman filter by evaluating the impact of a persistence state model, also known as a constant-state or random-walk model. In this model, the temperature at the next step is assumed to be equal to the previous temperature estimate, up to process noise:
\[
T_k = T_{k-1} + w_k,
\]
where \(w_k\) represents process noise. This assumption is appropriate when the underlying temperature changes gradually relative to the measurement rate, as is the case here, where we expect that at any given set of temperature measurements consecutive temperature states are expected to be similar. For the bias-augmented Kalman filter, the state vector was extended to include both temperature and model bias:
\[
\mathbf{x}_k =
\begin{bmatrix}
T_k \\
b_k
\end{bmatrix}.
\]
Both temperature and bias were modeled using the same persistence assumption:
\[
\mathbf{x}_k = \mathbf{x}_{k-1} + \mathbf{w}_k.
\]
The observation model was then written as
\[
z_k = T_k + b_k + v_k,
\]
where \(z_k\) is the native model prediction and \(v_k\) is measurement noise. This formulation allows the filter to smooth noisy model predictions while also estimating systematic bias in the native model outputs.

\begin{table}[htbp]
\centering
\caption{Comparison of model performance with and without Kalman filtering.  The PL RMSE is the testing error of each model before filtering.  The Kalman filter improves the estimates by leveraging the time continuity of the signal and is also able to correct the bias in the estimator.}
\label{tab:rmse_comparison}
\begin{tabular}{lcccc}
\hline
\textbf{Model} & \textbf{PL} & \textbf{KF} & \textbf{KF} & \textbf{Improvement} \\
\textbf{} & \textbf{RMSE ($^{\circ}$C)} & \textbf{(No Bias)} & \textbf{(With Bias)} & \textbf{with bias(\%)} \\
%\hline
Autoencoder & 1.66 & 1.55 & 1.56 & 6.8 \\
MLP & 0.20 & 0.16 & \textbf{0.15} & 23.1 \\
CNN & 0.53 & 0.48 & \textbf{0.22} & \textbf{60.4} \\
\hline
\end{tabular}
\end{table}

As shown in Table~\ref{tab:rmse_comparison}, the persistence-based Kalman filter reduced RMSE for all three models compared with the native predictions. The largest improvement was observed for the Convolutional Neural Networks (CNN) model, where the bias-augmented Kalman filter reduced RMSE from \(0.53\,^{ \circ}\mathrm{C}\) to \(0.22\, ^{\circ}\mathrm{C}\), corresponding to a \(60.4\%\) improvement. The Multi-Layer Perceptron (MLP) model also improved from \(0.20\,^{\circ}\mathrm{C}\) to \(0.15\,^{\circ}\mathrm{C}\). For the autoencoder, the no-bias Kalman filter gave a slight improvement over the native model, while adding bias estimation did not further improve performance. These results indicate that incorporating a temporal prior significantly improves the performance of data-driven models over short time periods without the need to incorporate sophisticated memory-dependent deep learning architectures such as Recurrent Neural Networks (RNN) and Long-Short Term Memory (LSTM) networks \cite{hochreiter1997long}.
% \paragraph{Summary of PL-ODMR fusion }
\par To improve the long-term stability of data-driven models, we implemented a hot-start strategy that fuses ODMR temperature estimates obtained from a physics-informed probabilistic model with  PL-measurement based temperature estimates from data driven models. %A regularized ReLU neural network was first trained to predict temperature from PL spectra. The validation RMSE of this PL model was squared and used as the PL measurement variance in the Kalman filter. 
As discussed above, at each temperature, the Kalman filter was reinitialized with the ODMR-predicted temperature estimate. The state uncertainty is set equal to the ODMR measurement variance obtained from the probabilistic model. Subsequently, the PL neural-network temperature estimates were then assimilated sequentially as noisy observations using a fixed process noise, until the next ODMR measurement became available and the filter was reinitialized again. Thus, ODMR measurement serves as an anchor, while PL provided fast, sequential temperature observations between the ODMR measurements. As shown in Fig \ref{fig:residuals} and Table \ref{tab:temperature_estimation_rmse} the KF approach reduces the testing errors by 57 \%. Principally, the gain in accuracy is due to the ability of the KF to reject outliers (Fig \ref{fig:histograms}). Our results demonstrate that KF approach can be exploited to maximally leverage the complementary strengths of PL (temporal resolution) and ODMR (long-term stability) measurements, laying out a viable path toward self‑correcting, high‑precision, field-deployable NV‑diamond temperature sensors.

\begin{table}[t]
\centering
\caption{Temperature-estimation performance for ODMR, PL, and KF fusion. RMSE values are reported in degrees Celsius.}
\label{tab:temperature_estimation_rmse}
\begin{tabular}{lccc}
\hline
\textbf{Evaluation set} & \textbf{ODMR } &  \textbf{PL} & \textbf{KF} \\
 & \textbf{RMSE ($^{\circ}$C)} &  \textbf{RMSE  ($^{\circ}$C)} & \textbf{RMSE  ($^{\circ}$C)} \\
Training & 0.78 & 0.16 & 0.082 \\
Testing & 0.95 & 0.94 & 0.54 \\
\hline
\end{tabular}
\end{table}
% Autoencoder & 1.6616 & 1.5486 & 1.5587 & 6.80 \\
% MLP & 0.2038 & 0.1635 & \textbf{0.1567} & 23.14 \\
% CNN & 0.5266 & 0.4758 & \textbf{0.2086} & \textbf{60.39} \\

%\section{Discussion}

\section{Summary}

We have investigated a Kalman‑filter‑based data‑fusion framework that combines the high‑speed, low‑CSWaP PL measurement-based temperature estimates arrived at using data driven-models together, with infrequent physics‑constrained optically detected magnetic resonance (ODMR) temperature estimates. By treating temperature as a slowly varying hidden variable and using a persistence (random‑walk) temporal prior on PL estimates, the filter smooths noisy PL‑derived predictions, reducing prediction errors by as much as 60\%. Furthermore, we extend the long-term temporal stability of the temperature model by exploiting a hot-start KF approach. Here, ODMR updates act as anchors that reset the filter state and correct systematic bias. Experimental validation over a temperature range of $15.0 ^ {\,\circ}\mathrm{C}$ to $45.0 ^{\,\circ}\mathrm{C}$  shows that the fused estimate achieves 57\% lower error than either modality alone, both for out‑of‑sample data. The improvement arises from the filter’s ability to reject outliers and allows us to maximally leverage the complementary strengths of PL and ODMR measurements. These results indicate that a KF based data fusion approach could dramatically improve the long-term performance of NV center based sensors, accelerating their development as fit-for purpose, field-deployable, robust sensors. 

\section{Acknowledgments}The authors thanks Adam Biacchi for providing the handheld spectrometer using in synchronous measurements. Tyrus Berry and Shraddha Rajpal acknowledge support from NIST Award 70NANB24H232 and Zeeshan Ahmed and Qiaochu Guo acknowledge support from AFMETCAL Award R24-685-0005.

%\section{Disclaimer}

%This document has not been peer reviewed but has been cleared by NIST for release.

\bibliographystyle{ieeetr}
\bibliography{references}

\end{document}